\documentclass[aps,pre, preprintnumbers ,floats,twocolumn,superscriptaddress]{revtex4}
\usepackage{graphicx}
\usepackage{dcolumn}
\usepackage{bm}
\usepackage[english]{babel}
\usepackage[latin1]{inputenc}
\usepackage{graphicx}
\usepackage{epstopdf}
\usepackage{amsfonts}
\usepackage{color}
\usepackage{epsfig}
\usepackage{mathrsfs}
\usepackage{amsmath,amssymb}
\usepackage{subfigure}
\usepackage{multirow}
\usepackage{url}
\usepackage{dsfont}

\usepackage{framed}
\usepackage[svgnames]{xcolor}

\begin{document}

\selectlanguage{english}


\title{Untangling the role of diverse social dimensions in the diffusion of microfinance}

\author{Elisa Omodei and Alex Arenas\\Department of Mathematics and Computer Science, Rovira i Virgili University}


\begin{abstract}
Ties between individuals on a social network can represent different dimensions of interactions, and the spreading of information and innovations on these networks could potentially be driven by some dimensions more than by others. In this paper we investigate this issue by studying the diffusion of microfinance within rural India villages and accounting for the whole multilayer structure of the underlying social networks. We define a new measure of node centrality, diffusion versatility, and show that this is a better predictor of microfinance participation rate than previously introduced measures defined on aggregated single-layer social networks. Moreover, we untangle the role played by each social dimension and find that the most prominent role is played by the nodes that are central on layers concerned with trust, shedding new light on the key triggers of the diffusion of microfinance.
\end{abstract}

\maketitle

\flushbottom


\section*{Introduction}

Understanding the mechanisms driving the diffusion of information, behaviours, and innovations is a question of great interest for social and economical sciences~\cite{coleman1957diffusion,rogers1962diffusion,bond201261}.
{\color{black}In his seminal book, Rogers identified four key elements for the diffusion of an innovation: the characteristics of the innovation itself, the communication channels, time, and the social systems within which the diffusion occurs~\cite{rogers1962diffusion}.
The role played by the social structure of the system has since then been widely investigated using the mathematical formalism of networks~\cite{valente1995network,watts2002simple}, and a fundamental question has been the identification of the most influential individuals therein~\cite{freeman1979centrality,kitsak2010identification}.
This is especially important in the context of \textit{network interventions}, which is concerned with understanding how social networks influence behaviours and their diffusions~\cite{valente2012network}.
In particular, induction interventions are designed to stimulate peer-to-peer interaction to trigger cascades in information or behavioural diffusion. 
Studies have shown that their success is critically dependent on the choice of influencers~\cite{valente1999accelerating} but also on their position in the network~\cite{aral2013engineering}.

In this paper, by taking advantage of the framework of multilayer networks, we investigate how the choice of opinion leaders can be improved in the context of the diffusion of microfinance in rural villages.}
Building on the seminal study of Banjeree \textit{et al.}~\cite{banerjee2013diffusion}, we rely on a unique dataset on social network structure and participation in microfinance of 43 villages in Karnataka, a state of southern India~\cite{data}. In-between 2007 and 2011 a microfinance institution, Bharatha Swamukti Samsthe (BSS), entered these villages, which previously had almost no exposure to any microfinance institution nor other types of formal credit. Before BSS's entrance in the villages, Banjeree and collaborators administered to households detailed surveys covering a wide range of interactions, to reconstruct the structure of the social network. When entering a village, BSS selected a number of pre-defined individuals that they would expect to be well connected within the villages (teachers, shopkeepers, leaders of self-help groups, etc), and had a private meeting with them to introduce the microfinance programme. These individuals, hereafter simply called leaders, then played a fundamental role in spreading the information about microcredit opportunities. Banerjee and collaborators investigate the correlation between the village level of participation in microfinance and the average centrality of its leaders in the social network.
{\color{black}
Their goal is to find the centrality measure that best predicts participation, so that in future interventions the most central individuals in the network could be selected as leaders to potentially maximise participation.
To the best of our knowledge, no study other than~\cite{banerjee2013diffusion} exists on applying the ideas of network interventions in the context of microfinance, the choice of the opinion leaders being left to credit institution criteria such as those just mentioned.
}

Banerjee and collaborators define, for each village, a social network of households as an undirected unweighted network linking two households if any of their members are in at least one of the relations covered by the survey.
They then introduce a new measure, called \textit{diffusion centrality}, to evaluate the importance of households within the network, with the ultimate goal of predicting the rate of village participation in microfinance on the basis of the centrality of the households that were firstly informed about it.
Given the network adjacency matrix $\mathbf{A}$, a passing probability $q$ and $T$ iterations, the diffusion centrality of node $i$ is the $i^{\text{th}}$ entry of the vector
\begin{equation}
DC(\mathbf{A};q,T) := \Big[ \sum_{t=1}^{T} (q\mathbf{A})^t \Big] \cdot \vec{\mathbf{1}}
\end{equation}
where $q$ is set as the inverse of the first eigenvalue of the adjacency matrix, and $T$ to the number of trimesters during which the village was exposed to BSS (6.6 on average). 
Essentially, diffusion centrality measures how effective a household would be as injection point of a new piece of information.
By means of multivariate linear regression (including 5 village-level controls, i.e. number of households, self-help group participation rate, savings participation rate, caste composition, and fraction of village households designated as leaders), they show that the average diffusion centrality of the pre-selected leaders outperforms other existing measures of centrality in predicting the village eventual rate of participation in microfinance.

The administered surveys, used to reconstruct the social network, cover 8 different dimensions: names of those whose homes the respondent visits or receives visits by, kins in the village, nonrelatives with whom the respondent socializes, those from whom the respondent receives medical help, those from which and to whom the respondent would borrow or lend money, those from which and to whom the respondent would borrow or lend material goods (such as kerosene or rice), those from or to whom the respondent gets or gives advice, and those with whom the respondent goes to pray (at a temple, church, or mosque).
{\color{black}In this paper, we show that taking into account the multilayer structure emerging from the different dimensions covered by the surveys leads to an improved prediction of microfinance participation. Moreover we investigate the relative role played by the different kinds of tie.
These results can be used in future network interventions in the context of microfinance, and beyond, to select opinion leaders in function of their position in the multilayer network, so to maximise participation in the programme.
The study is motivated by the recent growing literature on multiplex networks showing that taking into account the multilayer structure of social networks -- which consist of different kinds of ties, from kinship, to friendship and professional relations~\cite{wasserman1994social} -- can shed new light into its topological and dynamical properties~\cite{kivela2014multilayer}. Therefore in this paper we reconsider the question of how innovations diffuse by asking: do all kinds of tie play the same role or are some dimensions more influential than others in fostering the adoption of an innovation?}

\section*{Materials and Methods}

\subsection*{Data}
In each village, about half of the households completed surveys in which each member was asked to list the names of people in the village with whom they had a certain relationship. Households were selected through random sampling and stratification by religion and geographic sub-regions. For further information about data collection we refer the reader to the original paper~\cite{banerjee2013diffusion}, and the publicly available dataset~\cite{data}.
Individuals were asked the following questions:
\begin{enumerate}
\item Name the 4 non-relatives whom you speak to the most.  
\item In your free time, whose house do you visit?  
\item Who visits your house in his or her free time? 
\item If you needed to borrow kerosene or rice, to whom would you go? 
\item Who would come to you if he/she needed to borrow kerosene or rice? 
\item If you suddenly needed to borrow Rs. 50 for a day, whom would you ask?
\item Who do you trust enough that if he/she needed to borrow Rs. 50 for a day you would lend it to him/her? 
\item Who comes to you for advice? 
\item If you had to make a difficult personal decision, whom would you ask for advice? 
\item If you had a medical emergency and were alone at home whom would you ask for help in getting to a hospital? 
\item Name any close relatives, aside those in this household, who also live in this village. 
\item Do you visit temple/mosque/church? Do you go with anyone else? What are the names of these people?
\end{enumerate}

{\color{black} We observe that some pairs of questions are symmetric, as for instance "In your free time, whose house do you visit?" and "Who visits your house in his or her free time?".
The two questions, jointly considered, allow to reconstruct a network describing who visits whom within each village. The same stands for questions 4-5, 6-7 and 8-9, which allow to reconstruct, respectively, the network of potential material good loans, of potential money loans, and of advice relationships. Therefore, from the 12 questions we identify 8 different dimensions: nonrelative socialisation (1), house visits (2-3), material good potential loans (4-5), money potential loans (6-7), advice exchange (8-9), help in a medical emergency (10), kinship (11), and praying company (12).}

\subsection*{Methods}
\begin{figure*}[h!]
\begin{center}
\includegraphics[width=0.8\textwidth]{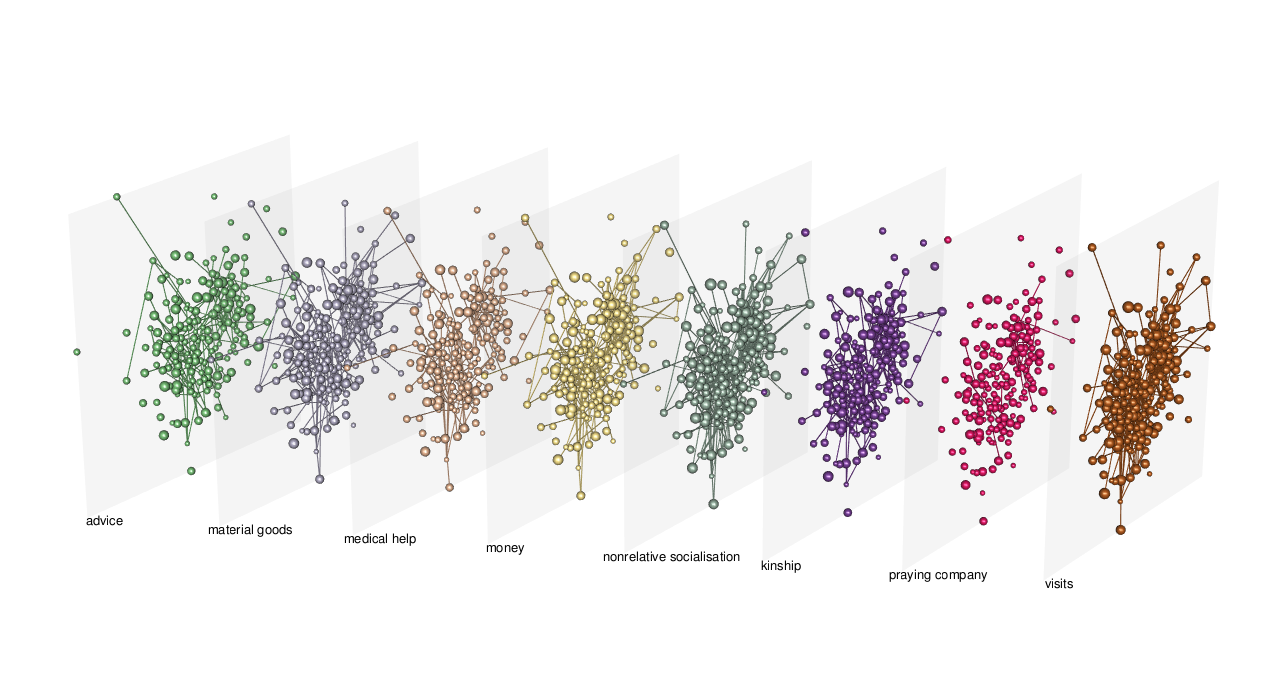}
\caption{\textbf{Multilayer social network.} Visualisation of the multilayer social network of one of the rural villages of southern India, obtained using the software MuxViz~\cite{de2014muxviz}. Each layer encodes a different dimension: visits, kinship, nonrelative socialisation, medical help, money, material goods, advice, and praying company. {\color{black}Nodes represent households and} their size is proportional to their layer-dependent diffusion versatility (Eq.~\ref{eq:diffverslayer}). Inter-layer links are omitted for a better rendering. {\color{black}Note that nodes have different connections on the different layers.}}
\label{multilayer}
\end{center}
\end{figure*}

The social network defined by Banerjee and collaborators is the product of an aggregation over different types of social ties, from kinship to medical help. 
It was recently shown that accounting for the whole multilayer structure of networks that are intrinsically composed of different kinds of relations has important consequences in the definition of the most central nodes, and allows to identify the more versatile ones~\cite{de2015ranking}. We call this extended notion of centrality \textit{versatility}. Here, we are interested in understanding if measuring leaders' versatility in a multilayer network that accounts for all dimensions separately can improve the prediction of microfinance participation.
To this end, for each village we build a multilayer network composed of $N$ nodes, corresponding to the number of households in the village, and $L=8$ layers, each encoding one of the dimensions defined above. Moreover, each node on a given layer is connected to its replica on all the other layers. Figure~\ref{multilayer} shows the visualisation of the multilayer social network for one of the villages.
Following the mathematical framework introduced in~\cite{de2013mathematical}, we describe this network by means of the rank-4 tensor {\color{black}$A^{\alpha\tilde{\gamma}}_{\beta\tilde{\delta}}$}. This was shown to be a natural generalisation of the adjacency matrix, and allows for a simple mathematical definition of multilayer networks{\color{black}, as we will now describe.

Let us first consider a standard network, composed of $N$ nodes and of only one single type of edge. Such graph can be represented by means of the adjacency matrix
\small
\begin{equation}
\mathbf{W}=\sum_{i,j=1}^{N}w_{ij}\mathbf{E}_{ij}=\sum_{i,j=1}^{N}w_{ij}\mathbf{e}_{i}\otimes \mathbf{e}_{j}^{\dag}\,, \quad \mathbf{A}\in\mathbb{R}^{N}\otimes\mathbb{R}^{N}=\mathbb{R}^{N\times N}\,,
\end{equation}
\normalsize
where $w_{ij}$ indicates the intensity of the relationship between node $i$ and node $j$, $\mathbf{e}_{i}$ is the canonical vector in the vector space $\mathbb{R}^{N}$, that is the $i^{\text{th}}$ component of $\mathbf{e}_{i}$ is 1, and all of its other components are $0$, and $\dag$ is the transposition operator, which transforms the column vector $\mathbf{e}_{j}$ into a row vector. $\mathbf{E}_{ij}=\mathbf{e}_{i}\otimes \mathbf{e}_{j}^{\dag}$ is the $2^{\text{nd}}$-order (i.e. rank-2) canonical tensor defined as the tensor product $\otimes$ of the two canonical vectors. 

Let us now introduce the language of tensors, that we need to generalise the notion of adjacency matrix to the more general notion of adjacency tensor needed to describe multilayer networks. 
We will use the covariant notation, in which a row vector $\mathbf{v}\in\mathbb{R}^{N}$ is given by a covariant vector $v_{\alpha}$ (where $\alpha=1,2,\ldots,N$), and the corresponding column vector $\mathbf{v}^{\dag}$ is given by the contravariant vector $v^{\alpha}$. Moreover, we will use Latin letters to denote the $i^{\text{th}}$ vector or the $(ij)^{\text{th}}$ tensor, and Greek letters to indicate the components of a vector or a tensor. Using this notation, $e_{\alpha}(i)$ is the $\alpha^{\text{th}}$ component of the $i^{\text{th}}$ covariant canonical vector $\mathbf{e}_{i}$ in $\mathbb{R}^{N}$, and $e^{\alpha}(j)$ is the $\alpha^{\text{th}}$ component of the $j^{\text{th}}$ contravariant canonical vector in $\mathbb{R}^{N}$. The adjacency matrix $\mathbf{W}$ can now be represented as rank-2 adjacency tensor $W^{\alpha}_{\beta}$ (1-covariant and 1-contravariant) as a linear combination of tensors in the canonical basis
\begin{equation}
W^{\alpha}_{\beta}=\sum_{i,j=1}^{N}w_{ij}e^{\alpha}(i)e_{\beta}(j)=\sum_{i,j=1}^{N}w_{ij}E^{\alpha}_{\beta}(ij)
\end{equation}
where $E^{\alpha}_{\beta}(ij)\in\mathbb{R}^{N\times N}$ indicates the tensor in the canonical basis corresponding to the tensor product of the canonical vectors assigned to nodes $i$ and $j$, i.e. it is $\mathbf{E}_{ij}$.

In a multilayer network, each type of relation between nodes is embedded in a different layer $\tilde{k}$ (where $\tilde{k}=1,2,\ldots,L$ and we use the tilda symbol to denote indices that correspond to layers).
For each of the layers, we construct the \emph{intra-layer} adjacency tensor $W^{\alpha}_{\beta}(\tilde{k})$ encoding information about relations between nodes within the same layer $\tilde{k}$. Moreover, to encode information about connections between nodes in different layers, we construct the \emph{inter-layer} adjacency tensors $C^{\alpha}_{\beta}(\tilde{h}\tilde{k})$. Note that, when $\tilde{h} = \tilde{k}$, we retrieve the intra-layer adjacency tensors $C^{\alpha}_{\beta}(\tilde{k}\tilde{k})=W^{\alpha}_{\beta}(\tilde{k})$.
Following the same approach as above, we define the covariant and contravariant vectors $e_{\tilde{\delta}}(\tilde{k})$ and $e^{\tilde{\gamma}}(\tilde{h})$ (where $\tilde{\delta}$, $\tilde{\gamma}$, $\tilde{k}$,$\tilde{h}$ all range in $(1,2,\ldots,L)$) of the canonical basis in the space $\mathbb{R}^{L}$. From these, we construct the $2^{\text{nd}}$-order tensors $E^{\tilde{\gamma}}_{\tilde{\delta}}(\tilde{h}\tilde{k})=e^{\tilde{\gamma}}(\tilde{h})e_{\tilde{\delta}}(\tilde{k})$ that represent the canonical basis of the space $\mathbb{R}^{L\times L}$.
Finally, we can now write the multilayer adjacency tensor as the tensor product between the adjacency tensors $C^{\alpha}_{\beta}(\tilde{h}\tilde{k})$ and the canonical tensors $E^{\tilde{\gamma}}_{\tilde{\delta}}(\tilde{h}\tilde{k})$: 
\begin{align}
\begin{split}
	A^{\alpha\tilde{\gamma}}_{\beta\tilde{\delta}} &= \sum_{\tilde{h},\tilde{k}=1}^{L}C^{\alpha}_{\beta}(\tilde{h}\tilde{k})E^{\tilde{\gamma}}_{\tilde{\delta}}(\tilde{h}\tilde{k})\\
	&= \sum_{\tilde{h},\tilde{k}=1}^{L} \Bigg[ \sum_{i,j=1}^{N}w_{ij}(\tilde{h}\tilde{k})E^{\alpha}_{\beta}(ij) \Bigg] E^{\tilde{\gamma}}_{\tilde{\delta}}(\tilde{h}\tilde{k})\\ 
	&= \sum_{\tilde{h},\tilde{k}=1}^{L}\sum_{i,j=1}^{N}w_{ij}(\tilde{h}\tilde{k})\mathcal{E}^{\alpha\tilde{\gamma}}_{\beta\tilde{\delta}}(ij\tilde{h}\tilde{k})
\end{split}
\end{align}
where $w_{ij}(\tilde{h}\tilde{k})$ are scalars that indicate the existence or not of a relationship between nodes $i$ and $j$, and $\mathcal{E}^{\alpha\tilde{\gamma}}_{\beta\tilde{\delta}}(ij\tilde{h}\tilde{k})\equiv e^{\alpha}(i)e_{\beta}(j)e^{\tilde{\gamma}}(\tilde{h})e_{\tilde{\delta}}(\tilde{k})$ is the $4^{\text{th}}$-order (i.e., rank-4) tensors of the canonical basis in the space $\mathbb{R}^{N\times N\times L\times L}$.

In our particular case, we define $w_{ij}(\tilde{h}\tilde{k})$ as follows. We set $w_{ij}(\tilde{k}\tilde{k}) = 1$ if there exists at least one member of household $i$ that indicated a relationship of type $\tilde{k}$ with any member of household $j$, or vice-versa, where $\tilde{k}$ refers to any of the socio-economic dimensions defined above. Moreover, to take into account the fact that the $L$ replicas of node $i$, one per layer, represent in fact the same household, we set $w_{ii}(\tilde{h}\tilde{k})=1$ for all $i=1,2,\ldots,N$ and all pairs of layers $(\tilde{h}\tilde{k})$. All others $w_{ij}(\tilde{h}\tilde{k})$ are set equal to 0.

We then generalise the definition of diffusion centrality by considering a diffusion process on the multilayer network, and introduce a new metrics that we call \textit{diffusion versatility}. We define the layer-dependent diffusion versatility of node $\alpha$ in layer $\tilde{\gamma}$ as the $(\alpha\tilde{\gamma})^{\text{th}}$ component of the rank-2 tensor
\begin{equation} \label{eq:diffverslayer}
DV_{\alpha\tilde{\gamma}}(A^{\alpha\tilde{\gamma}}_{\beta\tilde{\delta}};q,T) := \Big[ \sum_{t=1}^{T} q(A^t)^{\alpha\tilde{\gamma}}_{\beta\tilde{\delta}} \Big] u^{\beta\tilde{\delta}}
\end{equation} 
where $(A^t)^{\alpha\tilde{\gamma}}_{\beta\tilde{\delta}}$ is the $t$-th power of the rank-4 tensor, and $u^{\beta\tilde{\delta}}=\sum_{\tilde{h}=1}^{L}\sum_{i=1}^{N}e^\beta(i)e^{\tilde{\delta}}(\tilde{h})$ is the $N \times L$ rank-2 tensor with all components equal to 1. We then obtain the diffusion versatility of node $\alpha$ independently of the layer by contracting the index of the tensor with the contravariant vector $u^{\tilde{\gamma}}$ whose entries are all equal to 1, and normalising by dividing by $L$:
\begin{equation} \label{eq:diffvers}
DV_\alpha(A^{\alpha\tilde{\gamma}}_{\beta\tilde{\delta}};q,T) = \frac{1}{L} DV_{\alpha\tilde{\gamma}}(A^{\alpha\tilde{\gamma}}_{\beta\tilde{\delta}};q,T) u^{\tilde{\gamma}} \mbox{ .}
\end{equation}
}

Let us note that the layer-dependent diffusion versatility $DV_{\alpha\tilde{\gamma}}(A^{\alpha\tilde{\gamma}}_{\beta\tilde{\delta}};q,T)$ is not equivalent to computing diffusion centrality on a network composed only by layer $\alpha$, because here we are taking into account the whole multilayer network in its computation. Therefore diffusion versatility $DV_\alpha(A^{\alpha\tilde{\gamma}}_{\beta\tilde{\delta}};q,T)$ is not equivalent to computing diffusion centrality on the single layers separately and then taking their average for each node.

Conceptually, the diffusion versatility of a node measures how far a diffusion process starting on the node can spread on the multilayer network in a given amount of time $T$ (in our case, the number of trimesters during which the village was exposed to the microfinance institution). Accounting for the whole multilayer structure allows to capture along which kind of ties the diffusion is more likely to take place, and to assess whether the importance of nodes in the network as seeds of a diffusion process is more dependent on a dimension or another. For instance, a household that is very central in the aggregated network because it has several kinship ties with other households in the village, might have lower diffusion versatility in the multilayer network than another household that has the same centrality in the aggregated network but whose ties span over different dimensions because there live a very trusted person to whom people go to ask for advice, money and material goods.

\section*{Results}

\subsection*{Comparing centrality and versatility node rankings}

\begin{figure*}[h!]
\begin{center}
\includegraphics[width=0.8\textwidth]{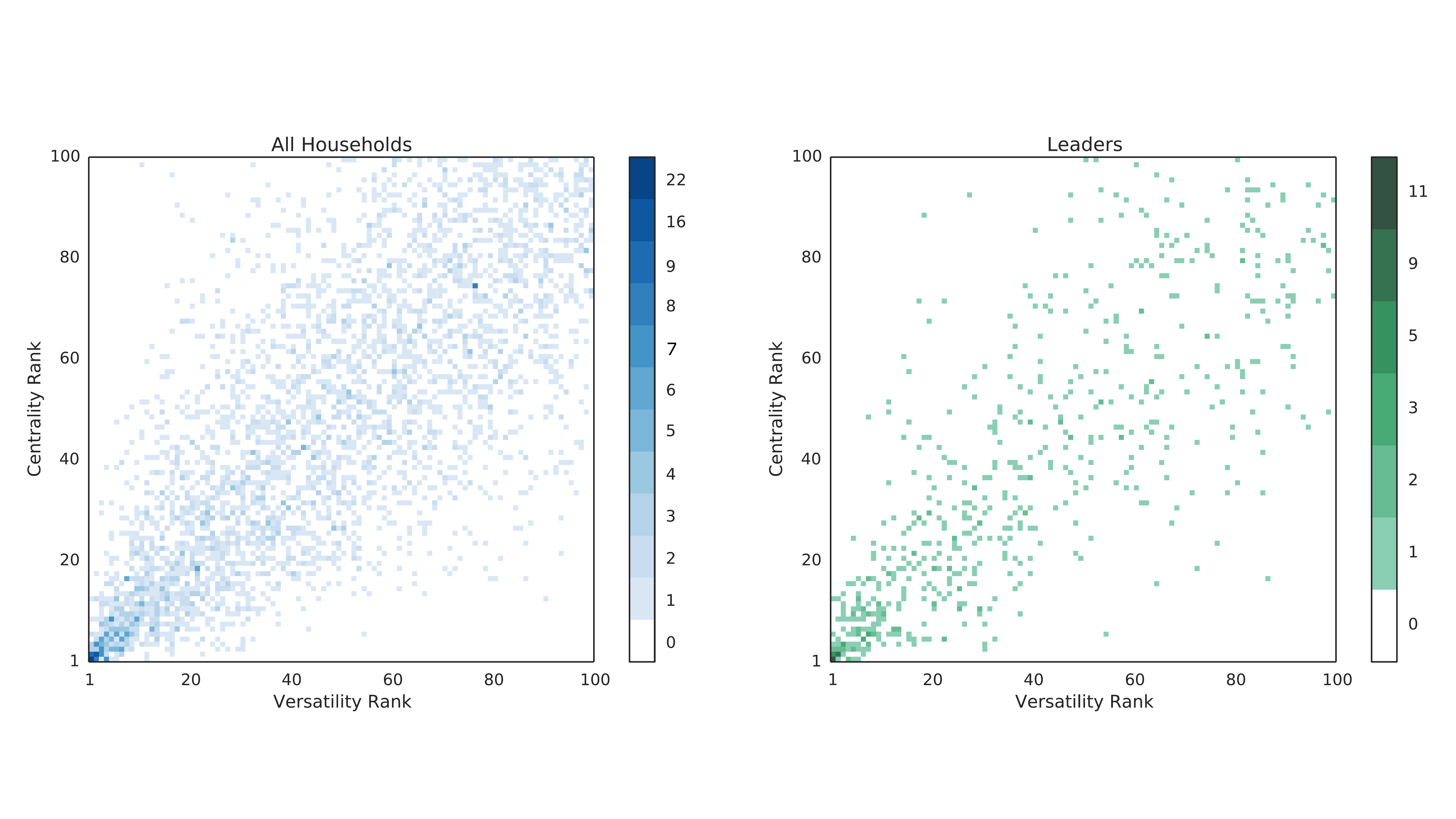}
\caption{\textbf{Node ranking comparison.} {\color{black}Density map} of node ranking according to their diffusion versatility and according to their diffusion centrality in the aggregated network, for the 100 top ranked nodes in each village {\color{black}(left panel), and for the subset of such nodes containing village leaders (right panel). We show that most nodes do not occupy the same position in the two rankings, with $28\%$ of them presenting a difference greater than or equal to 10, suggesting that diffusion versatility provides different information with respect to diffusion centrality.}}
\label{ranking}
\end{center}
\end{figure*}

First, we show that ranking nodes according to their diffusion versatility is significantly different than ranking them according to their diffusion centrality in the aggregated network. 
Figure~\ref{ranking} shows a {\color{black}density map} of the two rankings, for the 100 top ranked nodes in each village (i.e. about half of the nodes, on average). We selected the top 100 to avoid biases in the rank comparison due to the fact that pairs or groups of less central (or versatile) nodes might present the same value of centrality (or versatility) and therefore the same rank cardinal number, thus biasing the comparison between two different rankings.
We observe that the two rankings are positively correlated as expected (using the multilayer network structure should capture some different aspects but not drastically change the whole ranking), but also that indeed the ranking is significantly different for several nodes. More specifically, $96\%$ of the nodes do not occupy the same position in the two rankings, and $28\%$ of them present a rank difference greater than or equal to 10. 
This result suggests that diffusion versatility provides different information with respect to diffusion centrality, and in the following sections we explore whether this information can lead to a better prediction of microfinance participation, and, more importantly, to the detection of which kinds of tie play the most important role.

\subsection*{Improving microfinance participation prediction}

We investigate the correlation between the average diffusion versatility of leaders (as defined in Eq.~\ref{eq:diffvers}) and the rate of microfinance participation in the village, and compare the results with those obtained using diffusion centrality. As shown in Table~\ref{regression}, we find that diffusion versatility is more strongly correlated to microfinance participation rate than diffusion centrality ($R^2=0.470$ for versatility, versus $R^2=0.442$ for centrality). 
{\color{black}To test the significance of the difference between the two models, we generate 1000 bootstrapped samples of the data, perform the linear regressions on them, and then compare the two resulting distributions of the coefficient of determination using the paired samples t-test. We find that we can accept with a $99\%$ confidence level the alternative hypothesis that the the average $R^2$ of the model that uses versatility is higher than the average $R^2$ of the model that uses centrality.}
These results show that accounting for the whole multilevel structure of the different dimensions provides a better framework to identify the pre-defined set of leaders that microfinance agencies should initially inform in order to maximise participation.
However, given that the improvement in prediction is significant but relatively small, we are interested in understanding if some kinds of tie play a more fundamental role than others in the diffusion, and leaders should therefore be chosen according to their layer-dependent versatility in some particular layers.

\begin{table*}[h!]
\caption{\textbf{Microfinance participation versus centrality and versatility of leaders.} Values shown are coefficients from ordinary least-squares regression. Each column represents a different regression. The dependent variable is the microfinance participation rate of nonleader households in a village. The covariates are diffusion centrality (regression 1) and diffusion versatility (regression 2), averaged over the set of leaders, as well as 5 control variables: number of households, self-help group participation rate, savings participation rate, caste composition, and fraction of village households designated as leaders. Standard errors (in parenthesis) are robust to heteroskedasticity.}
\label{regression}
\begin{center}
\begin{tabular}{ l c c }
 & \multicolumn{2}{c}{\textbf{Regression}} \\
\textbf{Measure} & 1 & 2 \\
\hline
\multirow{ 2}{*}{diffusion centrality} & 0.022 (0.007) & \\
& P=0.002 & \\
\multirow{2}{*}{diffusion versatility} & & 0.030 (0.011) \\
& & P=0.001\\
$R^2$ & 0.442 & 0.470 \\
\hline
\end{tabular}
\end{center}
\end{table*}

\subsection*{Untangling the importance of the different dimensions}

We investigate whether the diverse dimensions contribute evenly, or rather play different roles, by considering the layer-dependent components of the diffusion versatility tensor, i.e. $DV_{\alpha\tilde{\gamma}}(A^{\alpha\tilde{\gamma}}_{\beta\tilde{\delta}};q,T)$. For each dimension, we compute the average leaders' versatility taking into account only the components of the corresponding layer $\tilde{\gamma}$, thus obtaining 8 different average leaders' versatilities, each corresponding to a given dimension. Let us note that this is not the same as computing diffusion centrality on each layer separately, because in this case each versatility value is computed taking into account the whole multilayer structure.
We perform 8 linear regressions, each using as covariate one of the 8 versatility measures (as well as the same control variables as above), and microfinance participation as the dependent variable. The results are reported in Table~2, from the least to the most predictive, as indicated by $R^2$ values. 
{\color{black}To assess the statistical significance of the difference between each of these models and the model based on diffusion centrality, we use paired samples t-test on 1000 bootstrapped samples of the data, as already described in the previous section. We find that we can accept with a $99\%$ confidence level the alternative hypothesis that the average $R^2$ of the the models that use layer-dependent versatility based on the layers \emph{material good}, \emph{kinship}, \emph{praying company}, \emph{advice}, \emph{money} and \emph{medical help} is higher than the average $R^2$ of the model that uses centrality. For the model based on the \emph{nonrelative socialisation} the confidence level is $90\%$. Instead, the average $R^2$ of the the model based on the \emph{visits} layer is smaller than the average $R^2$ of the model that uses centrality ($99\%$ confidence level). Moreover, we find that we can accept with a $99\%$ confidence level the alternative hypothesis that the average $R^2$ of the the models that use layer-dependent versatility based on the layers \emph{money} and \emph{medical help} is also higher than the average $R^2$ of the model that uses overall versatility. The same holds also for the \emph{advice} layer, but with a confidence level of $90\%$.}
These results indicate that the most predictive dimensions are all related to trust: asking for help in a medical emergency, asking for money if in need, and asking for advice. These results mean that the versatility of leaders in these layers is what best correlates with the final rate of participation in microfinance in the village. This could serve as an indication for microfinance institutions for leader selection, which could be done on the basis of diffusion versatility, but with a particular focus on individuals belonging to households which are particularly versatile on these specific layers.


\begin{table*}[h!]
\caption{\textbf{Microfinance participation versus layer-dependent versatility of leaders.} Values shown are coefficients from ordinary least-squares regression. Each column represents a different regression. The dependent variable is the microfinance participation rate of nonleader households in a village. The covariates are the layer-dependent diffusion versatility of the given layer, averaged over the set of leaders, as well as the 5 control variables.}
\begin{center}
\tiny
\begin{tabular}{ l c c c c c c c c }
 & \multicolumn{8}{c}{\textbf{Regression}} \\
\textbf{Dimension} & 1 & 2 & 3 & 4 & 5 & 6 & 7 & 8 \\
\hline
\multirow{ 2}{*}{visits} & 0.016(0.007) & & & & & & & \\
& P=0.019 & & & & & & & \\
nonrelative & & 0.021(0.007) & & & & & & \\
socialisation & & P=0.004  & & & & & & \\
\multirow{2}{*}{material goods} & & & 0.031(0.010) & & & & & \\
& & & P=0.004  & & & & & \\
\multirow{2}{*}{kinship} & & & & 0.034(0.010) & & & & \\
& & & & P=0.002  & & & & \\
praying & & & & & 0.044(0.013) & & & \\
company & & & & & P=0.002  & & & \\
\multirow{2}{*}{advice} & & & & & & 0.025(0.008) & & \\
& & & & & & P=0.003  & & \\
\multirow{2}{*}{money} & & & & & & & 0.028(0.009) & \\
& & & & & & & P=0.003 & \\
\multirow{2}{*}{medical help} & & & & & & & & 0.034(0.009) \\
& & & & & & & & P=0.000 \\
$R^2$ & 0.365 & 0.443 & 0.447 & 0.468 & 0.472 & 0.474 & 0.487 & 0.511 \\
\hline
\end{tabular}
\label{regression2}
\end{center}
\end{table*}

\section*{Conclusions}
In this paper we have shown that taking into account the multilayer structure of social networks of rural India villages allows for a better identification of the individuals who are more likely to help the spreading of microfinance in the community.
Firstly, we have introduced a new measure, diffusion versatility, as an extension of diffusion centrality to multilayer networks. We have shown that the diffusion versatility of leaders is a better predictor of the microfinance participation rate in the village than diffusion centrality.
Secondly, we have used the layer-dependent components of diffusion versatility to untangle the role played by each dimension in the diffusion of microfinance. 
We have found that the most predictive dimensions are related with trust: asking for help in a medical emergency or for a money loan if in need.

These results show that diffusion versatility could be used by microfinance institutions to identify opinion leaders so to maximise participation, focusing in particular on those with high versatility in specific layers.
Further field research could validate these results, for instance by means of randomised field experiments. Leaders in a set of villages could be chosen according to their layer-dependent diffusion versatility ranking relative to a given dimension, and in another set of villages according to a different dimension, and then compare participation.
Moreover, future work should involve sociologists and anthropologists in order to combine methods of multilayer network analysis with detailed investigations of the sociological meaning of the different dimensions in the context of rural India, to gain a deeper understanding of these social systems and how innovations diffuse therein.

\section*{Acknowledgements}
The authors would like to thank Matthew O. Jackson for the fruitful discussions. AA and EO were supported by the James S.\ McDonnell Foundation through grant 220020325. AA also acknowledges financial support from the European Commission FET-Proactive project MULTIPLEX (Grant No. 317532), the ICREA Academia and by Spanish government grant FIS2015-38266.
\small

\addcontentsline{toc}{section}{References}
\begin{small}
\bibliographystyle{apsrev4-1}

\end{small}

\end{document}